\documentclass{svjour3}                     
\smartqed  
\usepackage{graphicx}

\journalname{Journal of Low Temperature Physics - QFS2009}

\begin{document}

\title{Evidence for a Self-Bound Liquid State and the Commensurate-Incommensurate
Coexistence in 2D $^3$He on Graphite
}

\author{D. Sato \and D. Tsuji \and S. Takayoshi \and K. Obata
\and T. Matsui \and Hiroshi Fukuyama
}

\institute{
	Department of Physics, Graduate School of Science,
	The University of Tokyo, 7-3-1 Hongo, Bunkyo-ku, Tokyo 113-0033, Japan \\
	Tel.: +81-(0)3-5841-4193\\
	Fax: +81-(0)3-5804-4528\\
	\email{hiroshi@phys.s.u-tokyo.ac.jp}
}

\date{Received: date / Accepted: date}

\maketitle

\begin{abstract}
We made heat-capacity measurements of two dimensional (2D) $^3$He adsorbed on graphite
preplated with monolayer $^4$He in a wide temperature range (0.1 $\leq T \leq$
80 mK) at densities higher than that for the 4/7 phase ( = 6.8 nm$^{-2}$).
In the density range of 6.8 $\leq \rho \leq$ 8.1 nm$^{-2}$,
the 4/7 phase is stable against additional $^3$He atoms up to 20\%
and they are promoted into the third layer.
We found evidence that such promoted atoms form a self-bound 
2D Fermi liquid with an approximate density of 1 nm$^{-2}$ from the measured
density dependence of the $\gamma$-coefficient of heat capacity.
We also show evidence for the first-order transition between
the commensurate 4/7 phase and the ferromagnetic incommensurate phase
in the second layer in the density range of 8.1 $\leq \rho \leq$ 9.5 nm$^{-2}$.

\end{abstract}

\section{Introduction}
\label{intro}
Monolayer $^3$He adsorbed on a graphite substrate is an ideal model system for 
studies of strongly interacting two-dimensional (2D) Fermions.
In this system, we can vary areal density of Fermions over a wide range where 
various quantum phases 
are found \cite{GreywallPRB,Fukuyama_JPSJ}.
Especially, in the second layer, a commensurate (C) solid phase,
``the so called 4/7 phase", appears at the four sevenths density of that of
the first layer \cite{Fukuyama_JPSJ},
where the multiple spin exchange (MSE) interactions compete
each other yielding strong magnetic frustration.
It is believed that the magnetic ground state of this phase is the gapless
quantum spin-liquid (QSL) \cite{Ishida,Masutomi}. 
According to the first principles calculations based on the WKB approximation 
\cite{RogerMSE}, the three-particle ring exchange becomes dominant with increasing
density due to its less steric hindrance than the other exchange processes. 
Therefore, the magnetic ground state changes from the gapless QSL state to 
the frustrated ferromagnetic (FF) one with increasing density as was observed 
in the previous magnetization measurements in low fields \cite{Schiffer,Bauerle_thesis}.
Schiffer \textit{et al.} \cite{Schiffer} claimed the first-order transition between 
the two phases from the measured linear density dependence of $T =$ 0 
magnetization for the pure $^3$He system (hereafter, $^3$He/$^3$He/gr).
The neutron scattering experiment \cite{Lauter}
shows that, at high enough densities, the structure of the second layer is
the incommensurate (IC) solid with a triangular lattice.
However, the nature of the magnetic and structural transitions among those 
phases are not known in detail until now.

In this letter, we present results of heat-capacity measurements of 2D $^3$He
adsorbed on graphite preplated with monolayer $^4$He (hereafter, $^3$He/$^4$He/gr)
in a wide temperature range (0.1 $\leq$ $T$ $\leq$ 80 mK) at
higher densities beyond the 4/7 phase ($\rho \geq \rho_{4/7} = 6.8$ nm$^{-2}$).
The results show clearly how the QSL-FF transition evolves as a function
of increasing density.

\section{Experimental}
\label{exp}
The surface area ($A$) of Grafoil (exfoliated graphite) substrate used here is 556 m$^2$. 
The first layer $^4$He of 12.09 nm$^{-2}$ was prepared at $T = 4.2$ K, 
and then $^3$He overlayers were made at 2.7 K. 
We assume almost complete isotropic stratification between the first and 
second layers. 

The heat-capacity was measured with the relaxation method at
0.1 $\leq T \leq$ 1 mK and the adiabatic heat-pulse one at $T \geq 0.3$ mK.
The temperature of the samples was determined by a platinum-pulsed NMR
thermometer ($T \leq 30$ mK), and a carbon resistance one ($T \geq 20$ mK)
which are calibrated by a $^3$He melting curve thermometer.
Details of the experimental setup were described in the previous paper
\cite{MatsumotoPhysicaB}.

\begin{figure}[tbh]
 \begin{center}
  \begin{minipage}{0.65\hsize}
   \includegraphics[width=1\textwidth]{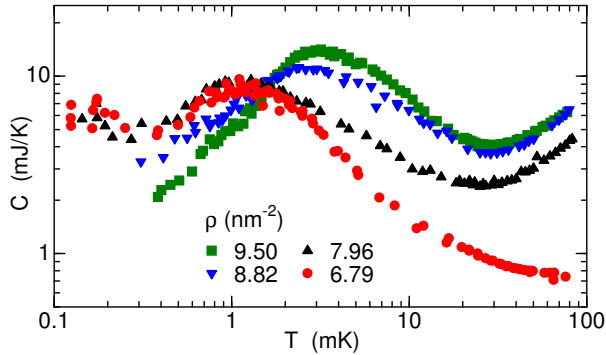}
  \end{minipage}
   \caption{(Color online) Heat capacities ($C$) of 2D $^3$He on graphite at four
   representative densities involving the commensurate 4/7 phase 
   ($\rho = 6.79$ and 7.96 nm$^{-2}$), the C-IC coexistence region
   (8.82 nm$^{-2}$), and the high density bound of the coexistence region 
   (9.50 nm$^{-2}$) for the second layer.
   	}
   \label{fig:1}      
 \end{center}
\end{figure}

\begin{figure}[htb]
 \begin{tabular}{cc}
  \begin{minipage}{0.48\hsize}
   \vspace*{0 cm}
   \includegraphics[width=1\textwidth]{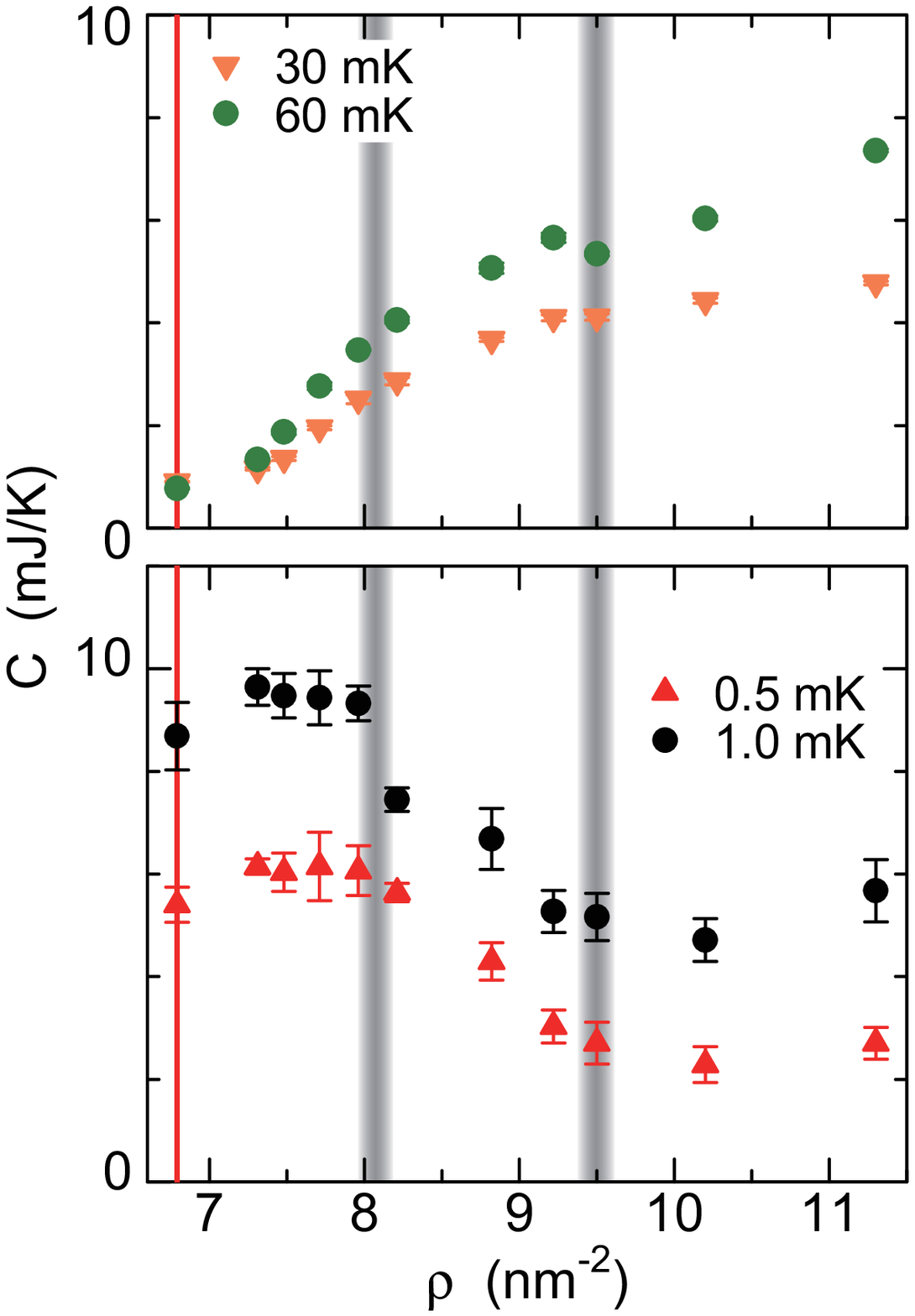}
   \end{minipage}
  \begin{minipage}{0.50\hsize}
   \includegraphics[width=1\textwidth]{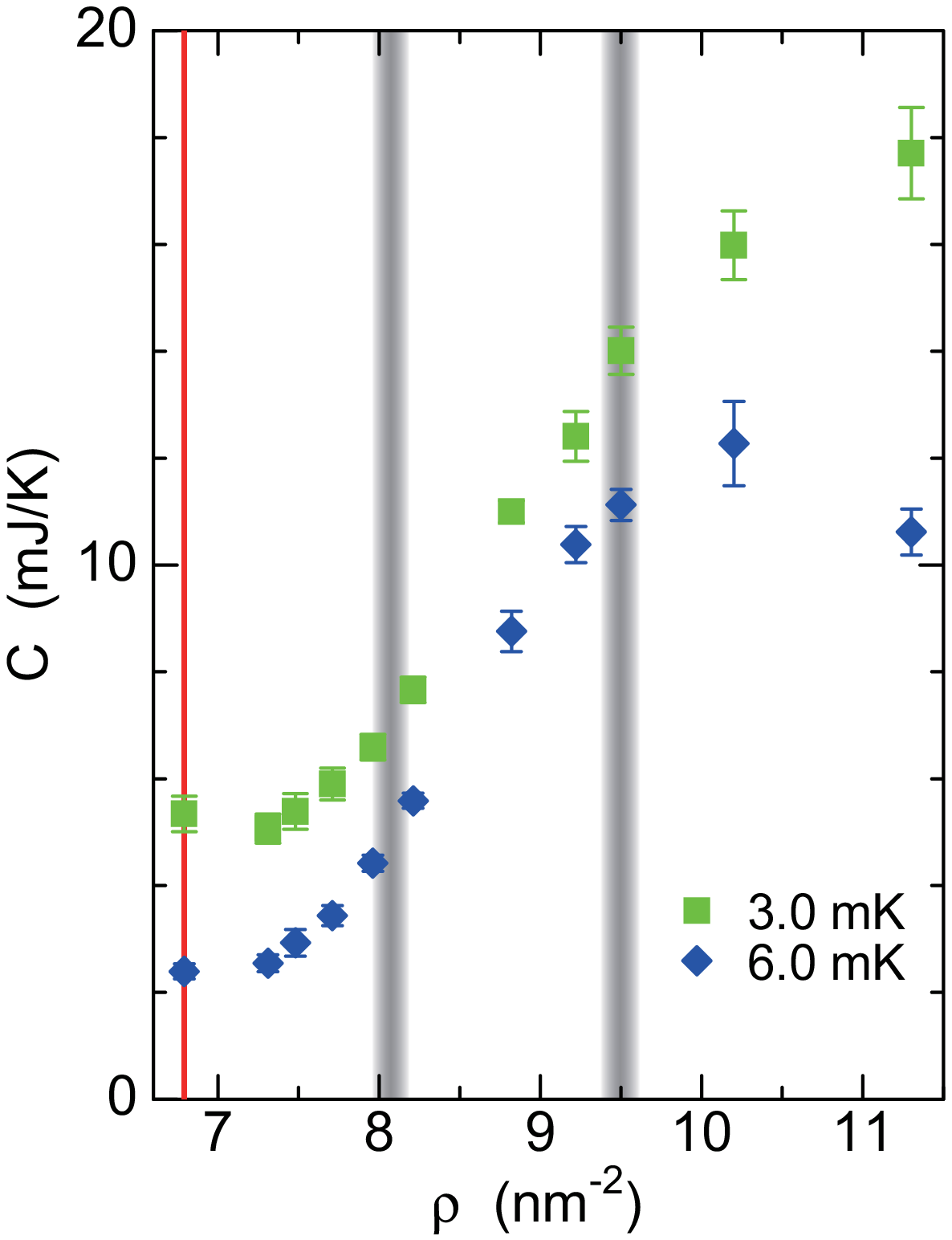}
  \end{minipage}
 \end{tabular}
   \caption{(Color online) Isotherms of $C$ at several temperatures.
	The vertical lines at $\rho = 6.8$ nm$^{-2}$ correspond to
	the 4/7 phase, and lines at $\rho = 8.1$ and 9.5 nm$^{-2}$
	are the low and high density bounds of the C-IC coexistence region.   	
   	}
   \label{fig:2}  
\end{figure}

\section{Results and Discussion}
\label{relsults}
Fig. \ref{fig:1} shows the measured heat capacities ($C$) at four 
representative densities.
The data can be classified into two distinct density regions. 
We discuss each region separately in the followings.

It is known that, in $^3$He/$^3$He/gr, the 4/7 phase ($\rho_{4/7} = 6.4$ nm$^{-2}$)
has two broad $C$ peaks at 1.8 and 0.3 mK, 
which is characteristic of highly frustrated magnetic systems \cite{Ishida}.
For the present system ($^3$He/$^4$He/gr), we also observed a similar
double-peak structure in the 4/7 phase but at 
somewhat lower temperatures, i.e., 1.2 and around 0.2 mK (see Fig. 1).
The peak-temperature differences are presumably caused by the different densities
of the 4/7 phase with respect to the first layer ones.

With increasing density, the peak temperatures are unchanged until $8.1 \pm 0.1$ nm$^{-2}$,
while the high-$T$ heat capacities increase progressively.
This can be seen more clearly in Fig. \ref{fig:2} where $C$ isotherms at several 
fixed temperatures are shown.
Excess heat capacities, $C_{FL} = C(\rho) - C(\rho_{4/7})$, can nicely be represented by the 
following functional form:
\begin{equation}\label{CFL}
C_{\rm FL} = \gamma T - \alpha T^2,
\end{equation}
which is characteristic of degenerate Fermi fluid. In 2D systems,
$\gamma = \pi k_B^2 m^{\ast} A / (3 \hbar^2)$, and $\gamma$ depends only on the $^3$He 
effective mass $m^{\ast}$ and $A$.
Therefore, we conclude that the 4/7 phase is stable up to $\rho \approx 8.1$ nm$^{-2}$ 
and the additional atoms are promoted into the third layer forming a degenerate
2D Fermi fluid in the temperature range we studied.

The $\gamma$-coefficients obtained in this way at densities below 10.3 nm$^{-2}$
are plotted as a function of the excess density ($\rho - \rho_{4/7}$) in Fig. 3.
The most striking feature here is that the $\gamma$ values are smaller 
than that for the ideal Fermi gas spreading over the whole surface ($\gamma_0$) with
$m^{\ast}/m =$ 1.
Moreover, $\gamma$ increases linearly with $\rho$ in the density region 
of 7.2 $\leq \rho \leq$ 8.1 nm$^{-2}$ and approaches $\gamma_0$
near 8.1 nm$^{-2}$.
This indicates that the third-layer fluid forms a self-bound 2D Fermi liquid (or $puddle$) 
with a fixed density around 1.3 nm$^{-2}$ covering a limited area of the surface.
The puddles cover the whole surface at $\rho \sim 8.1$ nm$^{-2}$ where 
the two phase coexistence between the C and IC solids starts (see below).
Although similarly small $\gamma$ values were noticed in the previous heat capacity
measurements in $^3$He/$^3$He/gr \cite{GreywallPRB,Siqueira}, they were 
considered as substrate heterogeneity effects not the puddle formation presumably 
because of the lack of data of the detailed density variations.
Also, the present observation is different from the previous suggestion 
by Godfrin {\it et al.}  \cite{Godfrin} who assumed the puddle formation 
of about 4 nm$^{-2}$ at much higher densities than we observed.

It has been believed that the critical point does not exist, i.e., the absence of the gas-liquid 
transition, in 2D $^3$He both theoretically \cite{Miller} and experimentally \cite{GreywallPRB}.
However, the variational Monte Carlo calculation taking account of delocalization 
of the wave function perpendicular to the 2D plane \cite{Brami} suggests 
the gas-liquid coexistence below about 2 nm$^{-2}$.
It seems to be plausible that the third-layer $^3$He on graphite can be self-bound 
due to the much smaller confinement potential than in the first- and second-layers.

Within a small excess-density regime below 0.5 nm$^{-2}$, the linear density dependence 
of $\gamma$ does not seem to hold (Fig. 3).
The most plausible explanation for this is that the first 7\% excess particles are 
somehow accommodated 
in the second layer and possibly in the third layer as well extending the area
of the 4/7 phase.
This explanation is consistent with small increases of the 0.5 and 1 mK isotherms
in the corresponding density region (see Fig. 2). 
If this is the case, the self-bound density should be 
about 1.0 rather than 1.3 nm$^{-2}$.
We don't know, at present, how the particles are accommodated in the commensurate
4/7 phase; compressing with domain walls? or filling intrinsically existing vacancies?

In Fig. 3 we also plotted the $\gamma$ values at higher densities than 8.1 nm$^{-2}$.
Here we fitted the raw $C$ data to the fitting function described 
below which is slightly different from eq. (1).
One can see a clear kink at $8.1 \pm 0.1$ nm$^{-2}$,
and the high density behavior seems to approach $\gamma_0$ at zero excess density.
This is a further support for the present scenario since the slow increase of $\gamma$ 
above 8.1 nm$^{-2}$ should be due to increasing $m^{\ast}$,
in the uniform 2D fluid.
Note that, in general, a commensurate phase (the 4/7 phase) can coexist only with a puddle phase not 
with a compressible fluid where the chemical potential varies with density.

\begin{figure}[tbh]
   \begin{center}
  \begin{minipage}{0.65\hsize}
   \includegraphics[width=1\textwidth]{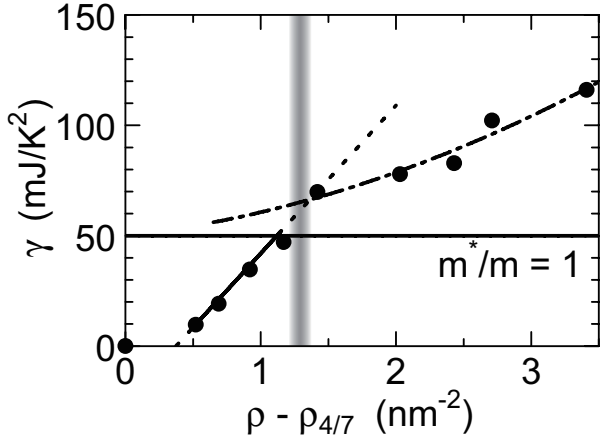}
  \end{minipage}
   \begin{minipage}{0.95\hsize}
   \caption{The $\gamma$-coefficient versus excess density from 
    the 4/7 phase ($\rho_{4/7} = 6.8$ nm$^{-2}$). The vertical line is the boundary between the commensurate
    region and two-phase coexistence region. The horizontal line
    corresponds to the $\gamma$ value of an ideal 2D Fermi gas ($m^{\ast}/m = 1$)
    for $A = 556$ m$^2$.
	The lines though the data points are guides to the eye.   	
   	}
   \label{fig:4}
  \end{minipage}
   \end{center}
\end{figure}

Above 8.1 nm$^{-2}$, the temperature dependence of $C$ dramatically changes. 
The 1 mK peak height rapidly decreases, while a new peak develops
rapidly near 3 mK.
This is clearly seen in Fig. 4 where the $C_{\rm FL}$ terms have already been subtracted.
We believe that this change is due to the magnetic QSL-FF transition associated with the
structural C-IC transition within the second layer and that the 3 mK peak 
corresponds to the FF IC phase \cite{Sato} from our previous magnetization 
measurements \cite{Murakawa}.
We fitted the data in this region to the following equation expected from the simple
coexistence model:
\begin{equation}\label{eqn2}
C = (1 - x) C_{\rm C} + x C_{\rm IC} + C_{\rm FL},
\end{equation}
where $C_{\rm C}$ and $C_{\rm IC}$ are smoothed heat-capacities measured 
at $\rho =$ 7.96 (C) and 9.50 nm$^{-2}$ (IC) without the Fermi fluid contributions 
$C_{\rm FL}$, respectively.
$x$ is the areal ratio of the IC to C solids in the second layer. 
The fittings are in good agreement with the experimental data as shown in Fig. 4,
and $x$ increases linearly with increasing $\rho$ (see the inset).
In addtion, all the data cross each other roughly at a single point (1.6 mK, 8.6 mJ/K) as expected.
Thus, the present results show unambiguously the first order transition between
the gapless QSL (C phase) and the FF state (IC phase) in the density region between
8.1 and 9.5 nm$^{-2}$.

The high-density bound of the coexistence ($\approx 9.5$ nm$^{-2}$) is determined as a density
above which the 3 mK peak starts to shift to lower temperatures.
This can be seen as a slope change at the corresponding  density in the $C$ isotherm at 0.5 mK  
in Fig. 2.
The shift is caused by lateral compression of the IC solid \cite{Sato}.
We note that the fittings to Eq. (2) give systematically larger heat capacities below 1 mK 
for the intermediate densities.
Particularly at 8.21 nm$^{-2}$, the fitting quality is worse, for which we don't know
the reason now.
Since low temperature magnetic properties of the 4/7 phase should be very sensitive even to 
small changes in competing interactions \cite{Fukuyama_JPSJ},
the discrepancy might be due to a possible interlayer interaction between the third-layer
fluid and the underlying 4/7 phase or to a domain wall structure in the C-IC transition.

\begin{figure}[tbh]
 \begin{center}
  \begin{minipage}{0.75\hsize}
   \includegraphics[width=1\textwidth]{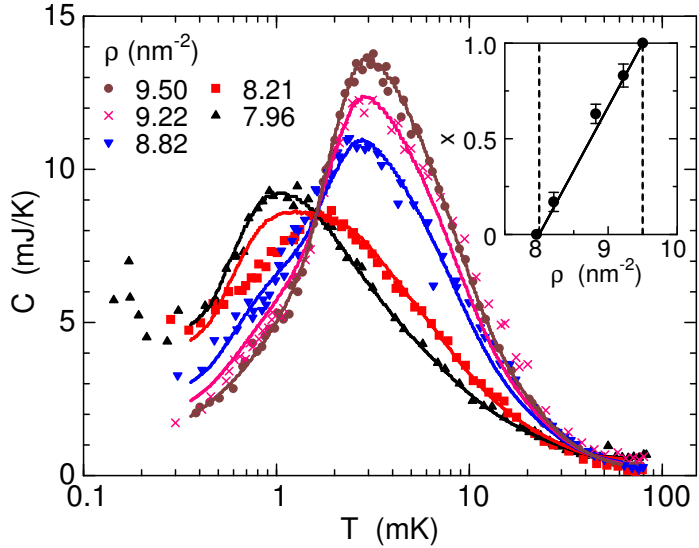}
  \end{minipage}

   \caption{(Color online) Heat capacities of the second layer solid $^3$He
   on graphite in the C-IC coexistence region. The solid lines are fittings
   to eq. (2) with $x$ = 0, 0.17, 0.63, 0.83, and 1 where 
   $x$ is the fraction of the IC solid. Note that the fluid overlayer
   contributions $C_{FL}$ are already subtracted here.
   The inset shows the density variation of $x$.
   }
   \label{fig:5}      
 \end{center}
\end{figure}

\section{Conclusion}
\label{conclusion}
From the heat capacity measurements down to 0.1 mK, we found that the commensurate 4/7 phase 
in the second layer on graphite is stable against additional particles up to 20\%
which are promoted into the third layer forming the low-density puddles of about 1 nm$^{-2}$.
We also obtained clear evidence for the first-order transition between the 4/7 phase with 
the gapless QSL ground state and the ferromagnetic incommensurate phase.

\begin{acknowledgements}
This work was financially supported by Grant-in-Aid for Scientific
Research on Priority Areas (No. 17071002) from MEXT, Japan.
D.S. acknowledges support from the Global COE Program ``the Physical Sciences Frontier", MEXT, Japan.
\end{acknowledgements}

\end{document}